\documentclass[prl,superscriptaddress,twocolumn]{revtex4}
\usepackage{amssymb}
\usepackage[tbtags]{amsmath}
\usepackage{graphicx}
\usepackage{epsfig,graphicx,times}

\begin{document}

\title{Controllable coherent population transfers in superconducting qubits for quantum computing}
\author{L.F. Wei}
\affiliation{CREST, Japan Science and Technology Agency (JST),
Kawaguchi, Saitama 332-0012, Japan}
\affiliation{Laboratory of Quantum Opt-electronics,
Southwest Jiaotong University,
Chengdu 610031, China}
\author{J.R. Johansson} \affiliation{Frontier Research System, The
Institute of Physical and Chemical Research (RIKEN), Wako-shi,
Saitama, 351-0198, Japan}
\author{L.X. Cen}
\affiliation{Department of Physics, Sichuan University, Chengdu,
610064, China}
\author{S. Ashhab}
\affiliation{Frontier Research System, The Institute of Physical
and Chemical Research (RIKEN), Wako-shi, Saitama, 351-0198, Japan}
\affiliation{Center for Theoretical Physics, Physics Department,
CSCS, The University of Michigan, Ann Arbor, Michigan 48109-1040,
USA}
\author{Franco Nori}
\affiliation{CREST, Japan Science and Technology Agency (JST),
Kawaguchi, Saitama 332-0012, Japan} \affiliation{Frontier Research
System, The Institute of Physical and Chemical Research (RIKEN),
Wako-shi, Saitama, 351-0198, Japan} \affiliation{Center for
Theoretical Physics, Physics Department, CSCS, The University of
Michigan, Ann Arbor, Michigan 48109-1040, USA}
\date{\today}

\begin{abstract}
We propose an approach to coherently transfer populations between
selected quantum states in one- and two-qubit systems by using
controllable Stark-chirped rapid adiabatic passages (SCRAPs).
These {\it evolution-time insensitive} transfers, assisted by
easily implementable single-qubit phase-shift operations, could
serve as elementary logic gates for quantum computing.
Specifically, this proposal could be conveniently demonstrated
with existing Josephson phase qubits. Our proposal can find an
immediate application in the readout of these qubits. Indeed, the
broken parity symmetries of the bound states in these artificial
``atoms" provide an efficient approach to design the required
adiabatic pulses.

PACS number(s):
42.50.Hz, 
03.67.Lx, 
85.25.Cp. 
\end{abstract}

\maketitle

{\it Introduction.---} The field of quantum computing is
attracting considerable experimental and theoretical attention.
Usually, elementary logic gates in quantum computing networks are
implemented using {\it precisely} designed resonant pulses. The
various fluctuations and operational imperfections that exist in
practice (e.g., the intensities of the applied pulses and
decoherence of the systems), however, limit these designs. For
example, the usual $\pi$-pulse driving for performing a
single-qubit NOT gate requires both a resonance condition {\it and
also} a {\it precise} value of the pulse area.  Also, the
difficulty of switching on/off interbit couplings~\cite{Wei-GHZ}
strongly limits the precise design of the required pulses for
two-qubit gates.

Here we propose an approach to coherently transfer the populations
of qubit states by using Stark-chirped rapid adiabatic passages
(SCRAPs)~\cite{NMR}. As in the case of geometric phases
\cite{Berry}, these population transfers are insensitive to the
dynamical evolution times of the qubits, as long as they are
adiabatic. Thus, here {\it it is not necessary to design
beforehand the exact durations of the applied pulses for these
transfers}. This is a convenient feature that could reduce the
sensitivity of the gate fidelities to certain types of
fluctuations. Another convenient feature of our proposal is that
the phase factors related to the transfer durations (which are
important for the operation of quantum gates) need only be known
{\it after the population transfer is completed}, at which time
they can be cancelled using easily implementable single-qubit
phase-shift operations. Therefore, depending on the nature of
fluctuations in the system, rapid adiabatic passages (RAPs) of
populations could offer an attractive approach to implementing
high-fidelity single-qubit NOT operations and two-qubit SWAP gates
for quantum computing. Also, the SCRAP-based quantum computation
proposed here is insensitive to the geometric properties of the
adiabatic passage paths. Thus, our approach for quantum computing
is distinctly different from both adiabatic quantum computation
(where the system is always kept in its ground state~\cite{Farhi})
and holonomic quantum computating (where implementations of
quantum gates are strongly related to the topological features of
either adiabatic or non-adiabatic evolution paths~\cite{lixiang}).

Although other adiabatic passage (AP) techniques, such as
stimulated Raman APs (STIRAPs)~\cite{raman}, have already been
proposed to implement quantum gates~\cite{shapiro07}, the present
SCRAP-based approach possesses certain advantages, such as: (i) it
advantageously utilizes dynamical Stark shifts induced by the
applied strong pulses (required to enforce adiabatic evolutions)
to produce the required detuning-chirps of the qubits, while in
STIRAP these shifts are unwanted and thus have to be overcome for
performing robust {\it resonant} drivings; and (ii) it couples
qubit levels directly via either one-photon or multiphoton
transitions, while in the STIRAP approach auxiliary levels are
required.

The key of SCRAP is how to produce time-dependent detunings by
chirping the qubit levels. For most natural atomic/molecular
systems, where each bound state possesses a definite parity, the
required detuning chirps could be achieved by making use of the
Stark effect (via either real, but relatively-weak, {\it
two-photon} excitations of the qubit levels~\cite{twophoton} or
certain virtual excitations to auxiliary bosonic
modes~\cite{virtual}). Here we show that the breaking of parity
symmetries in the bound states in current-biased Josephson
junctions (CBJJs) provides an advantage, because the desirable
detuning chirps can be produced by single-photon pulses. This is
because all the electric-dipole matrix elements could be nonzero
in such artificial ``atoms"~\cite{clarke88}. As a consequence, the
SCRAP-based quantum gates proposed here could be conveniently
demonstrated with driven Josephson phase qubits~\cite{berkely03}
generated by CBJJs. In order to stress the analogy with atomic
systems, we will refer to the energy shifts of the CBJJ energy
levels generated by external pulses as Stark shifts.

{\it Models.---} Usually, single-qubit gates are implemented by
using coherent Rabi oscillations. The Hamiltonian of such a driven
qubit reads $H_0(t)=\omega_0\sigma_z/2+R(t)\sigma_x$, with
$\omega_0$ being the eigenfrequency of the qubit and $R(t)$ the
controllable coupling between the qubit states; $\sigma_z$ and
$\sigma_x$ are Pauli operators. If the qubit is driven resonantly,
e.g., $R(t)=\Omega(t)\cos(\omega_0t)$, then the qubit undergoes a
rotation $R_x(t)=\cos[A(t)/2]-i\sigma_x\sin[A(t)/2]$, with
$A(t)=\int_0^t\Omega(t')dt'$. For realizing a single-qubit
NOT-gate, the pulse area is required to be {\it precisely}
designed as $A(t)=\pi$, since the population of the target logic
state $P(t)=[1-\cos A(t)]/2$ is very {\it sensitive} to the pulse
area $A(t)$ [in this example, we are assuming an initially empty
target state]. Relaxing such a rigorous condition, we additionally
chirp the qubit's eigenfrequency $\omega_0$ by introducing a
time-dependent Stark shift $\Delta(t)$. Therefore, the qubit
evolves under the time-dependent Hamiltonian
${H}'_0(t)=\omega_0\sigma_z/2+R(t)\sigma_x+\Delta(t)\sigma_z/2$,
which becomes
\begin{eqnarray}
H_1(t)=\frac{1}{2}\left(
\begin{array}{cc}
0&\Omega(t)\\
\Omega(t)&2\Delta(t)
\end{array}
\right)
\end{eqnarray}
in the interaction picture. Under the
condition
\begin{equation}
\frac{1}{2}\left|\Omega(t)\frac{d\Delta(t)}{dt}-\Delta(t)\frac{d\Omega(t)}{dt}\right|\ll[\Delta^2(t)+\Omega^2(t)]^{3/2},
\end{equation}
the driven qubit adiabatically evolves along two paths---the
instantaneous eigenstates
$|\lambda_{-}(t)\rangle=\cos[\theta(t)]|0\rangle-\sin[\theta(t)]|1\rangle$
and
$|\lambda_{+}(t)\rangle=\sin[\theta(t)]|0\rangle+\cos[\theta(t)]|1\rangle$,
respectively.
In principle, these adiabatic evolutions could produce arbitrary
single-qubit gates. For example, a single detuning pulse
$\Delta(t)$ (without a Rabi pulse) is sufficient to produce a
phase-shift gate: $U_z(\alpha)=\exp(i\alpha|1\rangle\langle 1|)$,
$\alpha=-\int_{-\infty}^{+\infty}\Delta(t)dt$. Furthermore,
combining the Rabi and detuning pulses for rotating the mixing
angle $\theta(t)=\arctan[\Omega(t)/\Delta(t)]/2$, from
$\theta(-\infty)=0$ to $\theta(+\infty)=\pi/2$, another
single-qubit gate
$U_x=\exp(i\beta_+)\sigma_+-\exp(i\beta_-)\sigma_-$ (with
$\beta_{\pm}=-\int_{-\infty}^{+\infty}\mu_{\pm}(t)dt,\,\mu_{\pm}(t)=\Delta(t)\pm\sqrt{\Delta^2(t)+\Omega^2(t)}$)
can be adiabatically implemented as:
\begin{eqnarray}
U_x:\left\{
\begin{array}{ll}
|\lambda_-(-\infty)\rangle=|0\rangle\overset{|\lambda_{-}(t)\rangle}{\longrightarrow}|\lambda_-(+\infty)\rangle=-e^{i\beta_-}|1\rangle,
\\
|\lambda_+(-\infty)\rangle=|1\rangle\overset{|\lambda_{+}(t)\rangle}{\longrightarrow}|\lambda_+(+\infty)\rangle=e^{i\beta_+}|0\rangle.
\end{array}
\right.
\end{eqnarray}
This is a single-qubit rotation that completely inverts the
populations of the qubit's logic states and thus is equivalent to
the single-qubit NOT gate. Note that here the population transfer
is {\it insensitive} to the pulse duration and other details of
the pulse shape---there is no need to precisely design these
beforehand. Different durations for finishing these transfers only
induce different additional phases $\beta_{\pm}$, which can then
be cancelled by properly applying the phase shift operations
$U_z(\alpha)$.

\begin{figure}[tbp]
\includegraphics[width=4.2cm, height=3cm]{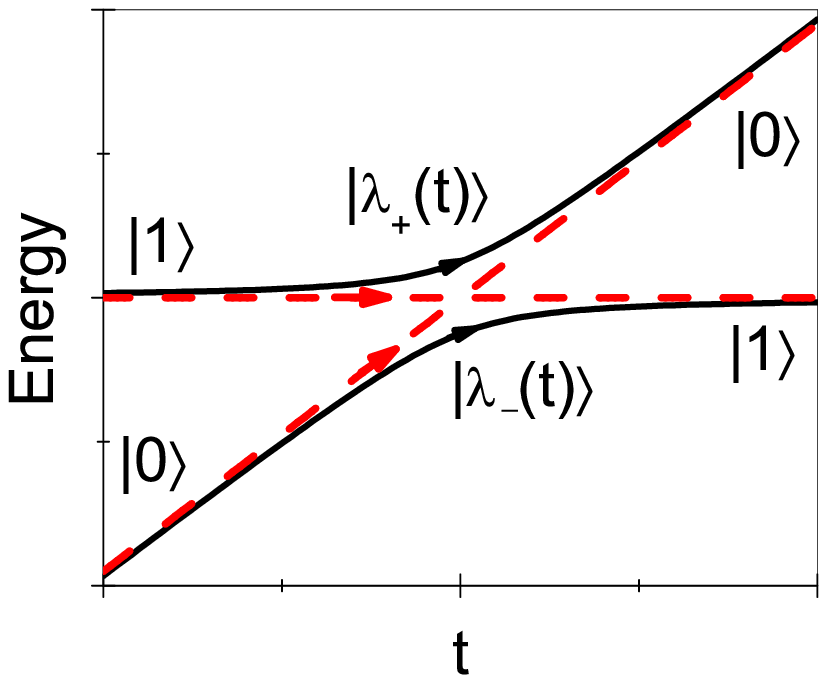}
\includegraphics[width=4.2cm, height=3cm]{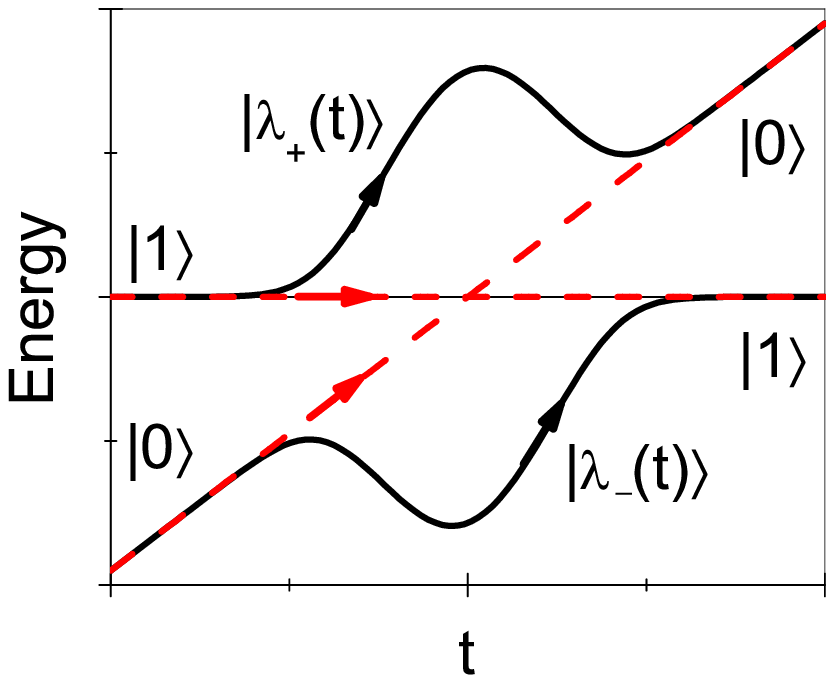}
\vspace{-0.5cm} \caption{(Color online) Simulated SCRAPs for
inverting the qubit's logic states by certain pulse combinations:
(left) a linear detuning pulse $\Delta(t)=v_at$, combined with a
constant Rabi pulse $\Omega(t)=\Omega_a$; and (right) a linear
detuning pulse $\Delta(t)=v_bt$, assisted by a Gaussian-shape Rabi
pulse $\Omega(t)=\Omega_b\exp(-t^2/T_R^2)$. Here, the solid
(black) lines are the expected adiabatic passage paths, and the
dashed (red) lines represent the unwanted Landau-Zener tunnelling
paths.}
\end{figure}

Similarly, the applied pulses are usually required to be {\it
exactly} designed for implementing two-qubit gates. For
example~\cite{xy-model}, for a typical two-qubit system described
by the XY-type Hamiltonian
$H_{12}=\sum_{i=1,2}\omega_i\sigma^{(i)}_{z}/2+K(t)\sum_{i\neq
j=1,2}\sigma^{(i)}_+\sigma^{(j)}_-/2$, with switchable real
interbit-coupling coefficient $K(t)$, the implementation of a
two-qubit SWAP gate requires that the interbit interaction time
$t$ should be {\it precisely} set as $\int_0^tK(t')dt'=\pi$ (when
$\omega_1=\omega_2$). This difficulty could be overcome by
introducing a time-dependent dc-driving to chirp the levels of one
qubit. In fact, we can add a Stark-shift term
$\Delta_2(t)\sigma^{(2)}_z/2$ applied to the second qubit and
evolve the system via (in the interaction picture)
\begin{eqnarray}
{H}_{2}(t)=\frac{1}{2}\left(
\begin{array}{cccc}
-\Delta_2(t)&0&0&0\\
0&-\Delta_2(t)&K(t)&0\\
0&K(t)&\Delta_2(t)&0\\
0&0&0&\Delta_2(t)
\end{array}
\right).
\end{eqnarray}
Obviously, three invariant subspaces; $\Re_0=\{|00\rangle\}$,
$\Re_1=\{|11\rangle\}$, and $\Re_2=\{|01\rangle,|10\rangle\}$
exist in the above driven dynamics. This implies that the
populations of states $|00\rangle$ and $|11\rangle$ are always
unchanged, while the evolution within the subspace $\Re_2$ is
determined by the reduced time-dependent Hamiltonian (1) with
$\Omega(t)$ and $\Delta(t)$ being replaced by $K(t)$ and
$\Delta_2(t)$, respectively.
Therefore, the APs determined by the Hamiltonian ${H}_2(t)$
produce an efficient two-qubit SWAP gate; the populations of
$|00\rangle$ and $|11\rangle$ remain unchanged, while the
populations of state $|10\rangle$ and $|01\rangle$ are exchanged.
The passages are just required to be adiabatic and again are {\it
insensitive} to the exact details of the applied pulses.

Figure 1 shows schematic diagrams of two simulated SCRAPs.
There, solid (black) lines are the desirable AP paths, and the
dashed (red) lines are the unwanted Landau-Zener
tunnellings~\cite{lz-book} (whose probabilities should be
negligible for the present adiabatic manipulations). These designs
could be similarly used to adiabatically invert the populations of
states $|10\rangle$ and $|01\rangle$ for implementing the
two-qubit SWAP gate.

{\it Demonstrations with driven Josephson phase qubits.---} In
principle, the above generic proposal could be experimentally
demonstrated with various physical systems~\cite{NMR}, e.g., the
{\it gas}-phase atoms and molecules, where SCRAPs are
experimentally feasible. Here, we propose a convenient
demonstration with {\it solid}-state Josephson junctions.

A CBJJ (see, e.g.,~\cite{berkely03}) biased by a time-independent
dc-current $I_b$ is described by
$\tilde{H}_0=p^2/2m+U(I_b,\delta)$.
Formally, such a CBJJ could be regarded as an artificial ``atom'',
with an effective mass $m=C_J\Phi_0/(2\pi)$, moving in a potential
$U(I_b,\delta)=-E_J(\cos\delta-I_b\delta/I_0)$. Here, $I_0$ and
$E_J=\Phi_0I_0/2\pi$ are, respectively, the critical current and
the Josephson energy of the junction of capacitance $C_J$. Under
proper dc-bias, e.g., $I_b\lesssim I_0$, the CBJJ has only a few
bound states: the lowest two levels, $|0\rangle$ and $|1\rangle$,
encode the qubit of eigenfrequency $\omega_{10}=(E_1-E_0)/\hbar$.
During the manipulations of the qubit, the third bound state
$|2\rangle$
of energy $E_2$ might be involved, as the difference between
$E_2-E_1$ and $E_1-E_0$ is relatively small.
Due to the broken mirror symmetry of the potential $U(\delta)$ for
$\delta\rightarrow-\delta$, bound states of this artificial
``atom'' {\it lose} their well-defined parities. As a consequence,
{\it all the electric-dipole matrix elements $\delta_{ij}=\langle
i|\delta|j\rangle, i,j=0,1,2 ,$ could be nonzero}~\cite{clarke88}.
This is essentially different from the situations in most natural
atoms/molecules, where {\it all} the bound states have {\it
well-defined parities} and the electric-dipole selection rule {\it
forbids} transitions between states with the same parity. By
making use of this property, Fig.~2 shows how to perform the
expected SCRAP with a single CBJJ by {\it only} applying an
amplitude-controlled dc-pulse $I_{\rm dc}(t)$ (to slowly chirp the
qubit's transition frequency) and a microwave pulse $I_{\rm
ac}(t)=A_{01}(t)\cos(\omega_{01}t)$ (to couple the qubit states).
Under these two pulses, the Hamiltonian of the driven CBJJ reads
$\tilde{H}_1(t)=\tilde{H_0}-(\Phi_0/2\pi)[I_{dc}(t)+I_{ac}(t)]\delta$.
Neglecting leakage, we then get the desirable Hamiltonian (1) with
$\Delta(t)=\tilde{\Delta}(t)=-(\Phi_0/2\pi)I_{dc}(t)(\delta_{11}-\delta_{00})$
and
$\Omega(t)=\tilde{\Omega}(t)=-(\Phi_0/2\pi)A_{01}(t)\delta_{01}$.
Obviously, for a natural atom/molecule with $\delta_{ii}=0$, the
present scheme for producing a Stark shift cannot be applied.

Specifically, for typical experimental parameters~\cite{berkely03}
($C_J=4.3\,$pF, $I_0=13.3\,\mu$A and $I_b=0.9725I_0$), our
numerical calculations show that the energy-splittings of the
lowest three bound states in this CBJJ to be $\omega_{10}=5.981$
GHz\,and $\omega_{21}=5.637$ GHz. The electric-dipole matrix
elements between these states are
$\delta_{00}=1.406,\,\delta_{11}=1.425,\,\delta_{22}=1.450,\,\delta_{01}=\delta_{10}=0.053,\,
\delta_{12}=\delta_{21}=0.077,$ and
$\delta_{02}=\delta_{20}=-0.004$. If the applied dc-pulse is a
linear function of time (i.e., $I_{\rm dc}(t)=v_1t$ with $v_1$
constant) and the coupling Rabi amplitude $\Omega(t)=\Omega_1$ is
fixed, then the above SCRAP reduces to the standard Landau-Zener
problem~\cite{lz-book}.
\begin{figure}[tbp]
\begin{minipage}[b]{3.00cm}
\includegraphics[height=2.0cm]{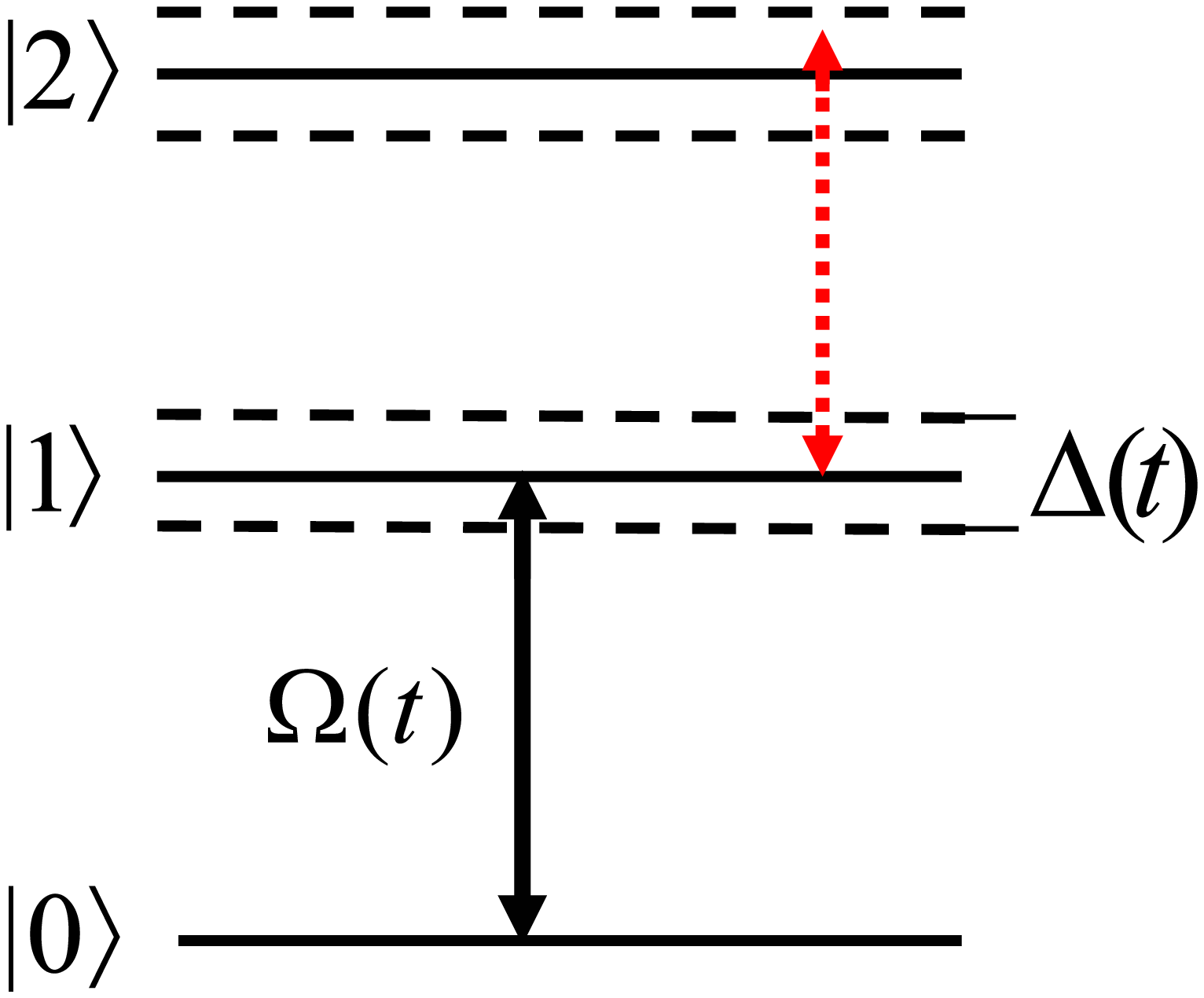}
\vspace{0.15cm}
\end{minipage}
\begin{minipage}[b]{4.50cm}
\includegraphics[width=4.0cm]{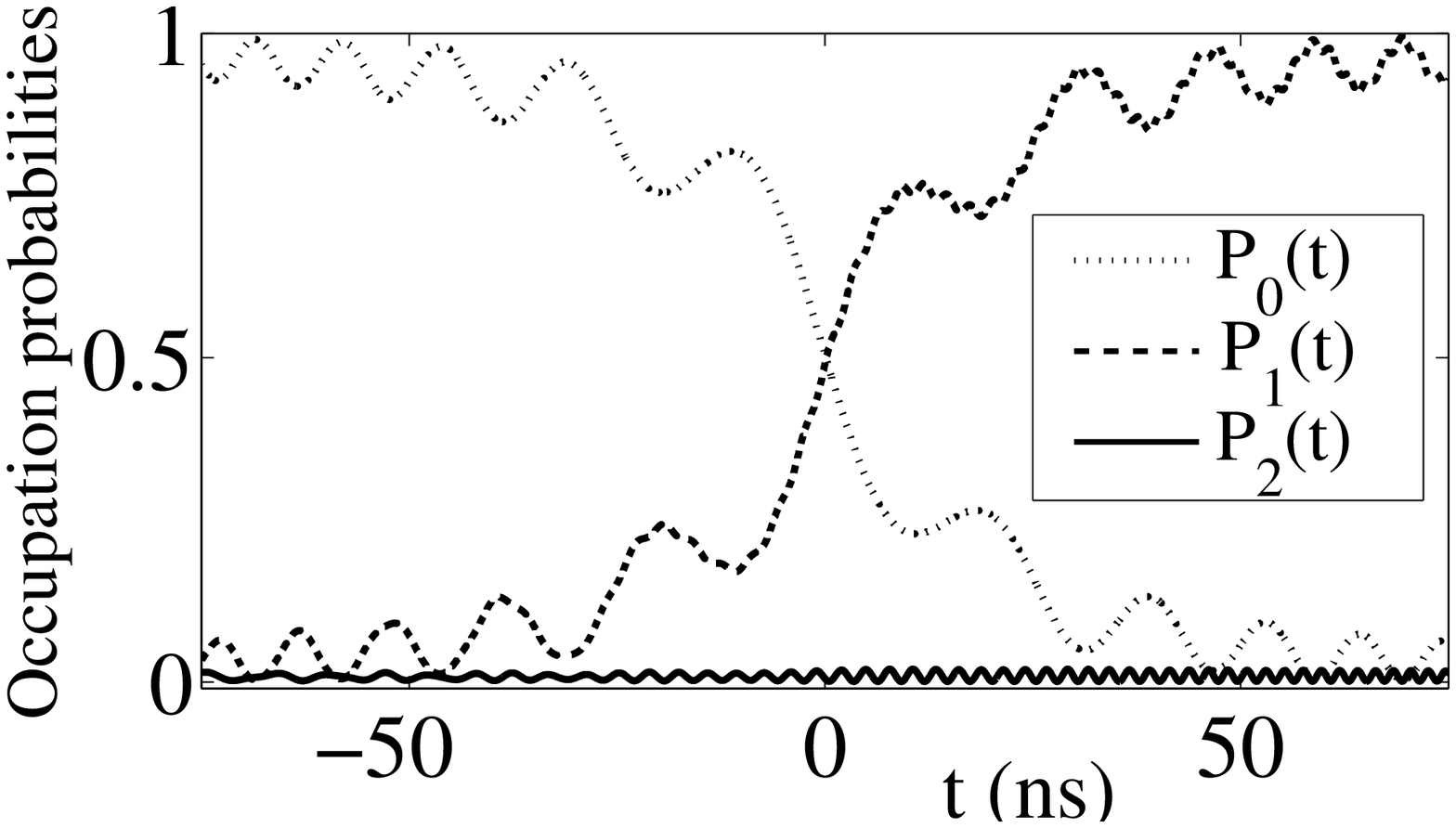}
\end{minipage}
\vspace{-0.2cm} \caption{(Color online) SCRAP-based population
transfers in a single Josephson phase qubit. (Left) manipulated
scheme: CBJJ levels with dashed chirped qubit energy splitting
$\Delta(t)$ is coupled (solid arrow) by a Rabi pulse $\Omega(t)$.
Dotted red arrow shows the unwanted leakage transition between the
chirping levels $|1\rangle$ and $|2\rangle$. (Right)
time-evolutions $P_j(t)$ of the occupation probabilities of the
lowest three levels $|j\rangle$ ($j=0,1,2$) in a CBJJ during the
designed SCRAPs for inverting the populations of the qubit logic
states. This shows that during the desirable SCRAPs the qubit
leakage is negligible.}
\end{figure}
For a typical driving with $v_1=0.15$ nA/ns and $A_{01}=1.25$ nA,
Fig.~2 simulates the time evolutions of the populations in this
three-level system during the designed SCRAPs. The unwanted (but
practically unavoidable) near-resonant transition between the
chirping levels $|1\rangle$ and $|2\rangle$ (due to the small
difference between $\omega_{21}$ and $\omega_{10}$) has been
considered. Figure 2 shows that during the above passages the
leakage to the third state $|2\rangle$ is sufficiently small.
Thus, the above proposal of performing the desirable SCRAPs to
implement single-qubit gates should be experimentally robust.

The adiabatic manipulations proposed above could also be utilized
to read out the qubits. In the usual readout
approach~\cite{berkely03}, the potential barrier is lowered fast
to enhance the tunneling and subsequent detection of the logic
state $|1\rangle$. Recently \cite{Lucero}, a $\pi$-pulse resonant
with the $|1\rangle\leftrightarrow |2\rangle$ transition was added
to the readout sequence for improved fidelity; The tunnelling rate
of the state $|2\rangle$ is significantly higher than those of the
qubit levels, and thus could be easily detected. The readout
scheme used in \cite{Lucero} can be improved further by utilizing
the above SCRAP by combining the applied microwave pulse and the
bias-current ramp. The population of state $|1\rangle$ is then
transferred to state $|2\rangle$ with very high fidelity. In
contrast to the above APs for quantum logic operations, here the
population transfer for readout is not bidirectional, as the
population of the target state $|2\rangle$ is initially empty. The
fidelity of such a readout could be very high, as long as the
relevant AP is sufficiently fast compared to the qubit decoherence
time.

Similarly, SCRAPs could also be used to implement two-qubit gates
in Josephson phase qubits. With no loss of generality, we consider
a superconducting circuit~\cite{berkely03} produced by
capacitively coupling two identical CBJJs. The SWAP gate is
typically performed by requiring that the two CBJJs be biased
identically (yielding the same level structures) and the static
interbit coupling between them reach the maximal value $K_0$. If
one waits precisely for an interaction time $\tau=\pi/2K_0$, then
a two-qubit SWAP gate is produced~\cite{delft07}. In order to
relax such {\it exact} constraints for the coupling procedure, we
propose adding a controllable dc current, $I_{\rm
dc}^{(2)}(t)=v_2t$, applied to the second CBJJ. Thus one can drive
the circuit under Hamiltonian
$\bar{H}_{12}(t)=\sum_{k=1,2}
H_{0k}+(2\pi/\Phi_0)^2p_1p_2/\bar{C}_m
-(\Phi_0/2\pi)I_{dc}^{(2)}(t)\delta_2$.
Here, the last term is the driving of the circuit, and the first
term $H_{0k}=(2\pi/\Phi_0)^2
p_k^2/(2\bar{C}_J)-E_{J}\cos\delta_k-(\Phi_0/2\pi)I_{b}\delta_k$
is the Hamiltonian of the $k$th CBJJ with a renormalized
junction-capacitance $\bar{C}_J=C_J(1+\zeta)$, with
$\zeta=C_m/(C_J+C_m)$. The coupling between these two CBJJs is
described by the second term with
$\bar{C}_m^{-1}=\zeta/[C_J(1+\zeta)]$ being the effective coupling
capacitance.
Suppose that the applied driving is not too strong, such that the
dynamics of each CBJJ is still safely limited within the subspace
$\emptyset_k=\{|0_k\rangle, |1_k\rangle, |2_k\rangle$\}:
$\sum_{l=0}^2|l_k\rangle\langle l_k|=1$. The circuit consequently
evolves within the total Hilbert space
$\emptyset=\emptyset_1\otimes\emptyset_2$.
Using the interaction picture defined by the unitary operator
$U_0=\prod_{k=1,2}\exp(-it\sum_{l=0}^2|l_k\rangle\langle l_k|)$,
we can easily check that, for the dynamics of the present circuit,
three invariant subspaces (relating to the computational basis)
exist: (i) $\Im_1=\{|00\rangle\}$ corresponding to the
sub-Hamiltonian $\bar{H}_1=E_{00}(t)|00\rangle\langle 00|$ with
$E_{00}(t)=-[\Phi_0/(2\pi)]I_{dc}^{(2)}(t)\delta_{00}
+(2\pi/\Phi_0)^2p_{00}^2/\bar{C}_{m}$, $p_{ll'}=\langle
l_k|p_k|l'_k\rangle$ and $\delta_{ll'}=\langle
l_k|\delta_k|l'_k\rangle$;
(ii) $\Im_2=\{|01\rangle,|10\rangle\}$ corresponding to the
sub-Hamiltonian $\bar{H}_2(t)$ taking the form of Eq.~(1) with
$\Omega(t)=\bar{\Omega}=2(2\pi/\Phi_0)^2p_{10}^2/\bar{C}_{m}$ and
$\Delta(t)=\bar{\Delta}(t)=[\phi_0/(2\pi)]I_{dc}^{(2)}(t)(\delta_{11}-\delta_{00})$;
and
(iii)
$\Im_3=\{|02\rangle=|a\rangle,|11\rangle=|b\rangle,|20\rangle=|c\rangle\}$
corresponding to
$$
\bar{H}_3(t)=\left(
\begin{array}{ccc}
E_a(t)&\Omega_{ab}e^{-it\vartheta}&\Omega_{ac}\\
\Omega_{ba}e^{it\vartheta}&E_b(t)&\Omega_{bc}e^{-it\vartheta}\\
\Omega_{ca}&\Omega_{cb}e^{it\vartheta}&E_{c}(t)\\
\end{array}
\right),
$$
with $E_a(t)=-[\Phi_0/(2\pi)]I_{dc}^{(2)}(t)\delta_{22}
+(2\pi/\Phi_0)^2p_{00}p_{22}/\bar{C}_{m}$,
$E_b(t)=-[\Phi_0/(2\pi)]I_{dc}^{(2)}(t)\delta_{11}
+(2\pi/\Phi_0)^2p_{11}^2/\bar{C}_{m}$,
$E_c(t)=-[\Phi_0/(2\pi)]I_{dc}^{(2)}(t)\delta_{00}
+(2\pi/\Phi_0)^2p_{22}p_{00}/\bar{C}_{m}$;
$\Omega_{ab}=\Omega_{ba}=\Omega_{bc}=\Omega_{cb}=(2\pi/\Phi_0)^2p_{01}p_{12}/\bar{C}_m$,
$\Omega_{ac}=\Omega_{ca}=(2\pi/\Phi_0)^2p_{02}^2/\bar{C}_m$, and
$\vartheta=\omega_{10}-\omega_{21}$.
Under the APs for exchanging the populations of the states
$|10\rangle$ and $|01\rangle$, we can easily see that the
population of $|00\rangle$ remains unchanged. Also, after the
desired APs, the population of the state $|11\rangle$ should also
be unchanged. Indeed, this is verified numerically in Fig.~3 for
the typical parameters $\zeta=0.05$ and $v_2=3.0$ nA/ns.
Therefore, the desirable two-qubit SWAP gate could also be
effectively produced by utilizing the proposed SCRAPs.

\begin{figure}[tbp]
\vspace{0.1cm}
\includegraphics[width=4.25cm]{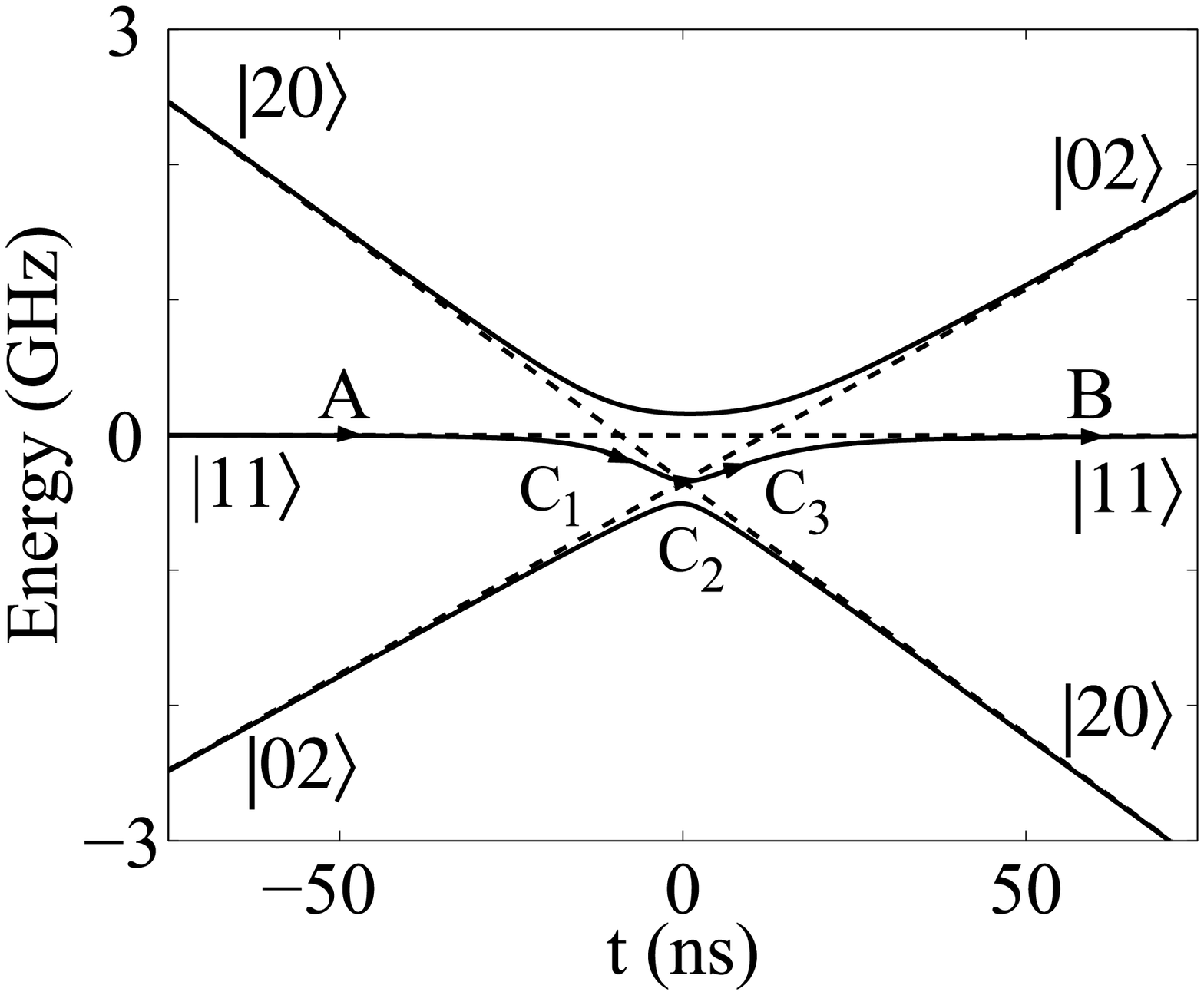}
\includegraphics[width=4.25cm]{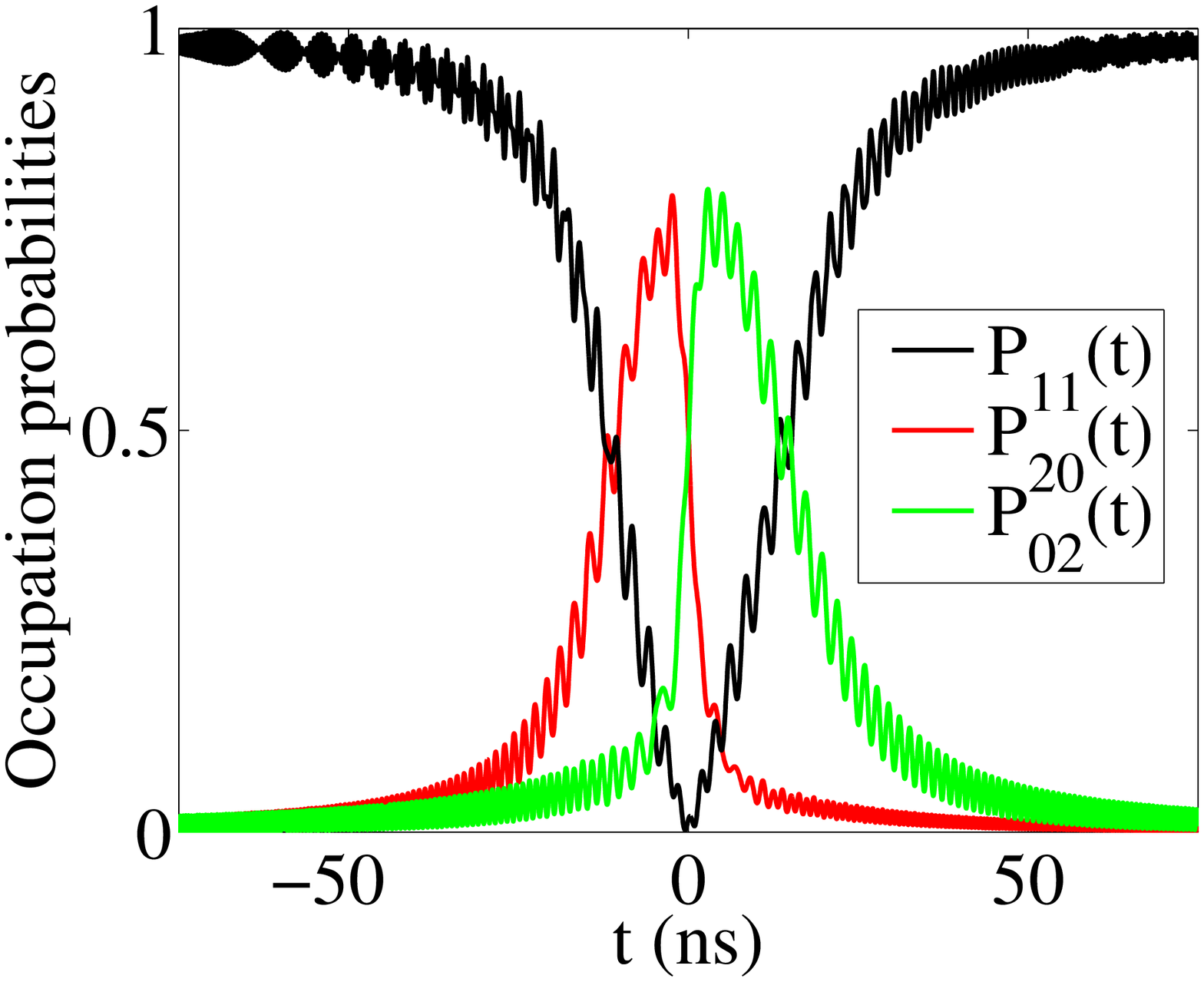}
\vspace{-0.2cm} \caption{(Color online) SCRAPs within the
invariant subspace $\Im_3=\{|02\rangle, |11\rangle, |20\rangle\}$
for the dynamics of two identical three-level capacitively-coupled
CBJJs driven by an amplitude-controllable dc-pulse. (Left)
Adiabatic energies and the desirable AP path (the middle
solid-line with arrows): $A\rightarrow C_1\rightarrow
C_2\rightarrow C_3\rightarrow B$. (Right) Time evolutions of
populations $P_{\alpha}(t),\,\alpha=20, 11, 02$, within the
invariant subspace $\Im_3$ during the designed SCRAPs for
inverting the populations of $|10\rangle$ and $|01\rangle$. It is
shown that the initial population of the $|11\rangle$ state
(corresponding to the $A$-regime in the left figure) is
adiabatically partly transferred to the two states $|20\rangle$
and $|02\rangle$ in the $C_1$, $C_2$, and $C_3$ regimes,
respectively. Note that the population of the state $|11\rangle$
vanishes at $t=0$ and completely returns after the passages.}
\end{figure}

{\it Discussions and Conclusions.---} By using SCRAPs, we have
shown that populations could be controllably transferred between
selected quantum states, insensitive to the details of the applied
adiabatic pulses. Assisted by readily implementable single-qubit
phase shift operations, these adiabatic population transfers could
be used to generate universal logic gates for quantum computing.
Experimentally existing superconducting circuits were treated as a
specific example to demonstrate the proposed approach.

Like other RAPs, the adiabatic nature of the present SCRAPs
requires that the passages should be sufficiently slow (compared
to the usual Rabi oscillations) and sufficiently fast (compared to
the decoherence times of the qubits). Satisfying both conditions
simultaneously does not pose any serious difficulty with typical
experimental parameters. Indeed, as we have shown above,
experimentally feasible APs could be applied within tens of
nanoseconds. This time interval is significantly longer than the
typical period of an experimental Rabi oscillation, which usually
does not exceed a few nanoseconds, and could be obviously shorter
than the typical decoherence times of existing qubits, which might
reach hundreds of nanoseconds, e.g., for the Josephson phase
qubits reported in ~\cite{berkely03}.
Solid-state qubits offer evident advantages due to their
scalability and controllability. Therefore, RAPs in solid-state
qubits could provide an attractive approach for data storage and
quantum information processing. We hope that such techniques will
be experimentally implemented in the near future.

This work was supported partly by the NSA, LPS, ARO, NSF grant No.
EIA-0130383; the National Nature Science Foundation of China
grants No. 60436010 and No. 10604043.

\end{document}